\documentclass{elsart3}
\usepackage{graphicx,amssymb,amsmath}
\usepackage[square,comma,sort&compress]{natbib}
\journal{International Conference on Magnetism}
\begin{document}
\begin{frontmatter}
\title{Clustered States as a New Paradigm of Condensed Matter Physics}
\author[nhmfl]{G. Alvarez}
and
\author[nhmfl]{E. Dagotto}
\address[nhmfl]{National High Magnetic Field Lab and Department of Physics,
Florida State University, Tallahassee, FL 32310}
\begin{abstract}
We argue that several materials of much current interest in condensed matter physics share
common phenomenological aspects that only very recent investigations are unveiling. The
list includes colossal magnetoresistance manganites, high temperature superconducting
cuprates, diluted magnetic semiconductors, and others. The common aspect is the relevance
of intrinsic inhomogeneities in the form of ``clustered states'', as explained in the text.
\end{abstract}
\end{frontmatter}

\section{Introduction}
In this paper it is argued that the recent trends unveiled 
in the context of manganites regarding the key role of
inhomogeneities to explain the colossal magnetoresistance effect
can be applied to a variety of other compounds as well.
Notably, evidence is rapidly accumulating that 
underdoped high temperature superconductors
are also inhomogeneous at the nanoscale, and theoretical approximations that
assume homogeneous states are questionable. It is still unclear if the
inhomogeneities have stripe features, as discussed extensively
in previous years, or whether they correspond
to more randomly shaped clusters, as assumed in the manganite 
context. Also the origin of these inhomogeneities is much debated. 
Computational simulations will play a key role in determining the properties
of models for manganites and cuprates, since percolative clustered physics is
difficult to handle with other methods. Even more recently,
clustered states have also been discussed in a very unexpected context,
the diluted magnetic semiconductors Mn-doped GaAs. These materials 
have considerable potential applications in spintronic devices and
understanding their behavior may raise their current critical temperatures.
In this publication, the three family of materials are briefly 
discussed and some key
references provided such that the reader can appreciate the common trends
between the many compounds. These quite unexpected similarities between
apparently very different materials promises to lead to a global unified view,
where clustered states become a new paradigm of condensed matter physics.

\section{Manganites}

\begin{figure}[h]
\centering{
\includegraphics[width=4.5cm]{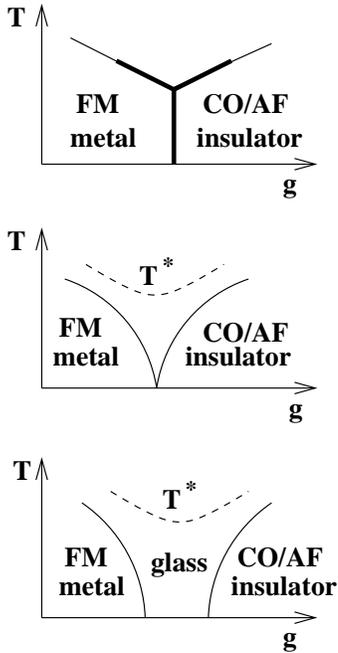}}
\caption{ {\it Top} General phase diagram of two competing phases 
in the absence of
quenched-disorder (or when this disorder is very weak). Thick (thin) lines
denote first (second) order transitions. 
$g$ is some parameter needed
to change from one phase to the other. \textit{Middle} With increasing
disorder, the temperature range with first-order transitions separating the
ordered states is reduced, and eventually for a \textit{fine-tuned} value of
the disorder the resulting phase diagram contains a quantum critical point.
In this context, this should be a rare occurrence. \textit{Bottom} In the
limit of substantial quenched
disorder, a window without any long-range order
opens at low temperature between the ordered phases.
This disordered state has glassy characteristics and it is composed of
coexisting clusters of both phases. The size of the coexisting islands is
regulated by the disorder strength and range, and by the proximity to the
original first-order transition. For more details 
see Refs.~\cite{burgy,book}. 
The new scale $T^{*}$, remnant of the
clean-limit transition, is also shown.}
\label{fig:tg3}
\end{figure}

\begin{figure}[h]
\centerline{\includegraphics[width=6cm]{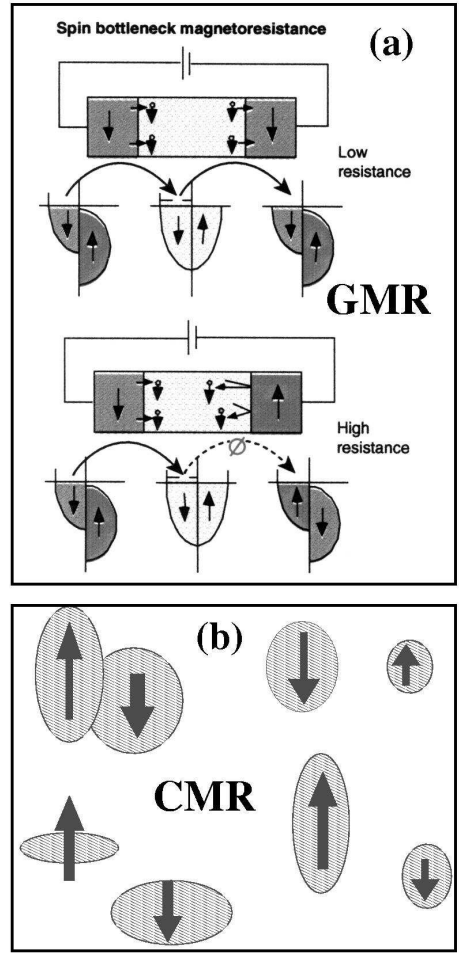}}
\caption{ Figure illustrating the analogies between CMR and GMR materials. 
(a) Schematic representation of a spin-valve effect. 
{\it (Top)} is made out of two ferromagnetic regions
with aligned spins. In this case low resistance is achieved. {\it
(Bottom)} consists
of antiparallel ferromagnetic moments, and in this case the device has a
high resistance. Reproduced from \cite{prinz}. (b) Ferromagnetic clusters in the CMR
materials. The arrows indicate orientation of magnetic moments (assuming Ising spins for
simplicity). Transport occurs if moments are aligned, as in GMR. Possible coexisting
insulating regions are not shown.
\label{fig:gmr}}
\end{figure}

Manganites are currently attracting considerable attention
mainly due to the presence of the colossal magnetoresistance effect
in magnetotransport measurements\cite{review,book}. 
In addition, these materials
have a complex phase diagram with a plethora of ordered phases,
a characteristic of correlated electron systems. A variety of experimental
and theoretical investigations
have unveiled the inhomogeneous character of the states of relevance
to explain the CMR phenomenon, with a competition between ferromagnetic
and antiferromagnetic states that induces coexistence of clusters
at the nanometer-scale \cite{book}.

To understand this
phenomenon, and its associated CMR effect, the
first-order transitions that separate the metallic and insulating phases
in the clean limit (i.e. without disorder) play a key role 
\cite{book,review}. The first-order character
of the transition is caused by the different magnetic and charge
orders of the competing states. The clean-limit 
phase diagram  is illustrated in 
Fig.~\ref{fig:tg3} (upper panel). When quenched
disorder is introduced in the coupling or density that is modified
to change from one phase to the other, the temperature where the N\'eel
and Curie temperatures meet is reduced in value and eventually collapses
to zero as in Fig.~\ref{fig:tg3} (middle panel). Further increasing
the disorder strength, a spin glassy
disordered region appears at low temperatures
involving coexisting clusters (Fig.~\ref{fig:tg3} (lower panel)).
Simulations by Burgy {\it et al.} \cite{burgy} have shown that the
clustered state between the Curie temperature and the clean-limit
critical temperature ($T^*$), with preformed 
ferromagnetic regions of random orientations,
has a huge magnetoresistance since {\it small fields can easily align
the moments of the ferromagnetic islands}, leading 
to a percolative conductor in agreement with experiments\cite{fath}.
This establishes qualitative similarities with ``Giant MR'' (GMR) multilayered materials, as
sketched in Fig.\ref{fig:gmr}.
The quenched disorder simply triggers the stabilization of the cluster
formation, but phase competition is the main driving force of the mixed state.
Elastic deformations may also play a key  role in this context\cite{bishop}.
We refer the reader to the literature cited here, and references therein,
for a more detailed view of this exciting area of investigations.
The readers can also consult a
 list of ``open problems'' in the manganite context that has been recently
provided\cite{recent}.

\section{High temperature superconductors}
\begin{figure}[h]
\centering{
\includegraphics[width=6.0cm]{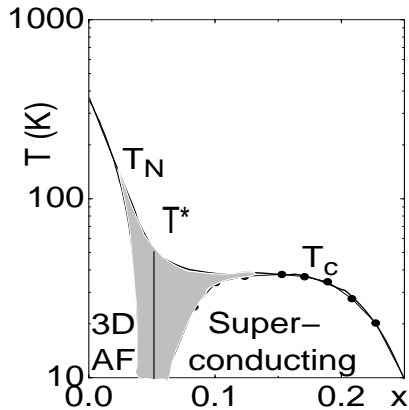}}
\caption{Conjectured HTS phase diagram. The black lines should be the actual phase
boundaries without disorder. The shaded region is conjectured to have metallic (SC) and
insulating (AF) coexisting regions in the real materials.
\label{fig:burgy}}
\end{figure}

The results obtained for manganites are not crucially
dependent on the nature of the two competing phases.
The argumentation should be valid in the context of high
temperature superconductors (HTS) as well, where an antiferromagnetic (AF)
insulator (probably with stripes) competes with a $d$-wave 
superconductor (dSC). In the clean-limit, difficult to obtain with
the standard chemical doping, the \emph{conjecture} is that the two
phases would have been separated by a line of first-order transitions
as shown in Fig.~\ref{fig:burgy}. The inevitable disorder of the chemical doping procedure
opens a window, and in this
context the glassy state that separates the insulator from the
superconductor is made up of small clusters of the two competing
phases\cite{burgy,kivelson} very different from the homogeneous exotic states proposed for that
regime.

First-order transitions AF-dSC have already appeared
in electron doped materials\cite{harima}, and they are also common
in heavy-fermion and inorganic superconductor contexts.

Considering the general character of the discussion here, there
should be an analog of the colossal effects in cuprates as well. In
fact,
{\it colossal effects should be ubiquitous when ordered phases compete}.
A possible manifestation of this effect could be the ``Giant
Proximity Effect'' recently discussed \cite{decca}
with  uncorrelated preformed SC clusters -- in the sense that their
phases in the order parameter are random -- 
percolating under the influence of nearby superconducting materials
and leading to a strong Josephson coupling across the originally
non-superconducting sample. 

These conjectured results for cuprates, where inhomogeneities play a key role,
are in excellent agreement with the most exciting recent experimental
developments in the high-Tc arena, namely the scanning tunneling
microscopy experiments that unveiled the clustered nature of these
materials at the nanoscale.\cite{davis}

\section{Diluted Magnetic Semiconductors}

Diluted magnetic semiconductors (DMS) are attracting
much attention lately due to their potential for device applications in the
growing field of spintronics. In particular, a large number of
DMS studies have focused on III-V compounds where Mn doping in InAs
and GaAs has been achieved using molecular beam epitaxy (MBE) techniques.
The main result of recent experimental efforts
is the discovery\cite{ohno1,katsumoto,potashnik}
of ferromagnetism at a Curie temperature $T_{\rm C}$$\sim$110~K in
Ga$_{1-x}$Mn$_x$As, with Mn concentrations $x$ up to 10\%. It is widely
believed that this ferromagnetism is ``carrier induced'', with holes
donated by Mn ions mediating a ferromagnetic interaction between
the randomly localized Mn$^{2+}$ spins. In practice,
anti-site defects reduce the number of holes $n$ from its ideal
value $n$=$x$, leading to a ratio $p$=$(n/x)$ substantially smaller than one.

\begin{figure}[h]
\centering{
\includegraphics[width=6.3cm]{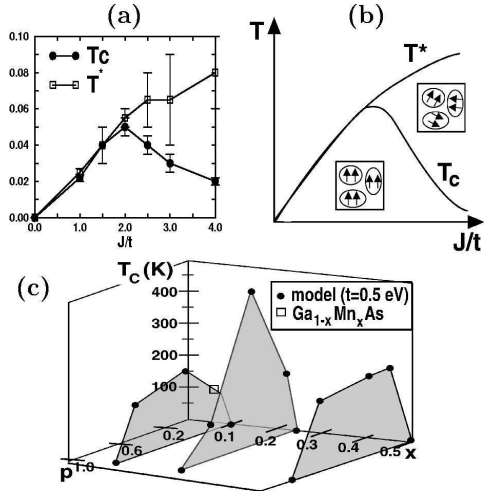}}
\caption{(a) Schematic phase diagram varying 
$J/t$, at fixed $x$ and $p$. At large $J/t$, 
a broad scale $T^*$ corresponds to the 
formation of uncorrelated clusters, as indicated.
$T_{\rm C}$ is the ``true''
transition temperature. At small $J/t$, those temperatures are similar. The
optimal $J/t$ is intermediate between itinerant and
localized regimes. (b) is the same as (a), but indicating the clustered states and moment
orientations.
(c) Numerically obtained $T_{\rm C}$ vs. $x$ and $p$, 
at $J/t$=2.0. Filled
circles are from model Eq.(1) 
with $t$=0.5 eV, while the open square corresponds to the
experimental value for Ga$_{1-x}$Mn$_x$As at $x$$\sim$0.1.
(Results from Ref.~\cite{paper1})
\label{fig:paper1}}
\end{figure}

\begin{figure}[h]
\centering{
\includegraphics[width=6.0cm]{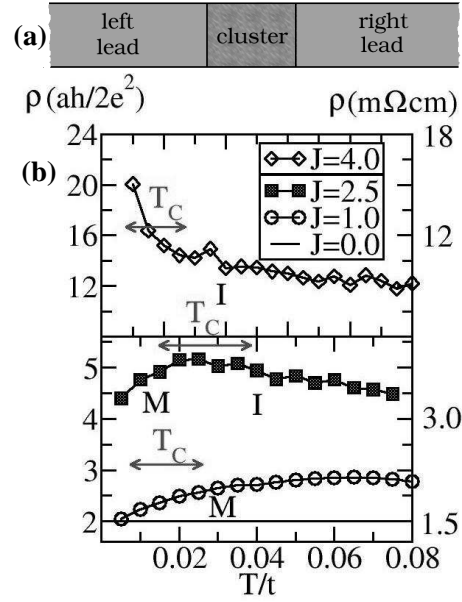}}
\caption{
(a) Geometry used for the calculation of the conductance. 
The interacting region (cluster) is connected by
ideal contacts to semi-infinite ideal leads.
(b) Dependence of the theoretically calculated resistivity, $\rho$,
with temperature in three dimensions.
Shown is  $\rho=L/G$ vs. $T$ on $L^3=4^3$ lattices, 16 spins ($x=0.25$), and 5 carriers
($p=0.3$) for the $J/t$'s indicated. An average over 20 disorder configurations has been
performed in each case. Units are shown in two scales, $a h/(2e^2)$ on the left and 
m$\Omega$cm on the right, with $L=4$ and assuming \mbox{$a=5.6$ \AA}.
 The estimated critical temperatures are also shown (the arrows indicate the current accuracy of the estimations).
(Results from Ref.~\cite{paper2})\label{fig:rho4d3}}
\end{figure} 
In this section, generic Monte Carlo (MC) studies of a diluted Kondo lattice 
model for DMS are summarized. \cite{paper1,paper0,paper2} Further information can be obtained from the cited
references. The Hamiltonian is similar to those employed in the Mn-oxide context, namely
\begin{equation}
{\hat H}=-t\sum_{<ij>,\sigma}{\hat c}^\dagger_{i\sigma} {\hat c}_{j\sigma} +
J\sum_{I}\vec{S}_I\cdot\vec{\sigma}_I,
\label{eq:ham}
\end{equation}
\noindent where 
${\hat c}^\dagger_{i\sigma}$ creates a hole at site $i$
with spin $\sigma$, and the hole spin operator interacting
antiferromagnetically with the localized Mn-spin $\vec{S}_I$ is
$\vec{\sigma}_I={\hat
c}^\dagger_{I\alpha}\vec{\sigma}_{\alpha,\beta}{\hat c}_{I\beta}$.
The carrier can visit $any$ site of the lattice.
The interaction term is restricted to randomly selected 
sites, $I$, with a $S$=5/2  Mn-moment. 
These spins are here considered classical with $|\vec{S}_I|$=1, as widely
assumed, allowing for a MC simulation technically similarly as 
in manganites\cite{review,book}. 
Approximations include the neglect
of on-site 
Coulomb repulsion $U$, valid at small $x$ and $p$ where double occupancy
is unlikely even at $J/t$=0. In addition, 
Mn-oxides investigations \cite{book} show that an intermediate or large
$J/t$ plays a role analogous to a Hubbard $U/t$ at {\it any} 
$x$ \cite{review}. 
At low $x$, the probability of nearest-neighbors (NN) Mn-spins is also 
low (0.0625 at $x$=0.25), justifying the neglect of an
anti-ferromagnetic (AF) Mn-Mn coupling. The 
hole motion is described 
by a one-band tight-binding model, while many bands
may be necessary \cite{macdonald}. Despite this simplification, our model
 contains spins and holes in interaction
and it is 
expected to capture the main qualitative aspects of carrier-induced 
ferromagnetism in DMS materials. 
Our effort builds upon previous
important DMS theoretical studies
\cite{theory,paper0}. However, it
 differs from previous
approaches in important qualitative aspects as explained in Ref.~\cite{paper1}.

One of the most important observations in the study of Eq. (\ref{eq:ham}) reported in
\cite{paper1}, 
is the non-trivial dependence of $T_{\rm C}$ with
coupling,    
$J/t$. 
For $J/t$$\rightarrow$$\infty$ and a Mn dilute system, the holes are trapped
in Mn-sites, 
reducing drastically $T_{\rm C}$. 
Small FM clusters of spins are formed 
at a temperature scale $T^*$, 
but there is no correlation between them, leading to a global vanishing
magnetization \cite{paper0}. On the other hand, when $J/t=0$ the system is
non-interacting.   
Since both in the  $J/t$$\sim$0 and 
$J/t$=$\infty$ limits $T_{\rm C}$ is suppressed, an
{\it optimal} $J/t|_{opt}$ must exist where $T_{\rm C}$ is
maximized. This can be seen
 in Fig.\ref{fig:paper1}a-b.\cite{paper1} 

The ($x$,$p$) dependence of $T_{\rm C}$ is shown in Fig.~\ref{fig:paper1}c, assuming 
$t=0.3$eV.\cite{paper1}
In this figure, the best value of $T_{\rm C}$ achieved experimentally for GaMnAs is also indicated 
approximately.   
Note that increasing $x$ beyond the experimental value of 0.1 would significantly increase
$T_{\rm C}$. At $J/t|_{opt}$, the best
value is $x$$\sim$0.25, but room-$T$ ferromagnetism would be possible even
with $x$$\sim$0.15. An increase of $T_{\rm C}$ could also be achieved if 
the compensation could be decreased; in the optimal case the best value would be close to
 $p=0.5$. These predictions seem in agreement with experimental developments
since very recently 
Ga$_{1-x}$Mn$_x$As samples with $T_{\rm C}$ as high as $150$ K were 
prepared,\cite{ku} a result believed to be caused by an enhanced free-hole density.

Dynamical and transport properties have also been calculated for the model presented
here \cite{paper2}. Of special importance is the conductance, $G$, which is calculated  
using the Kubo formula adapted to geometries usually employed in the
context of mesoscopic systems.\cite{verges}  The cluster is considered 
to be connected by ideal contacts to 
two semi-infinite ideal leads, as represented in Fig.~\ref{fig:rho4d3}a. In 
Fig.~\ref{fig:rho4d3}b 
the inverse of the conductance, which is a measure of the resistivity, 
is plotted for a
three-dimensional lattice at weak, intermediate, and strong coupling,
at fixed $x$=$0.25$ and $p$=$0.3$. For the weak 
coupling regime ($J/t$=$1.0$) the system is weakly metallic 
at all temperatures. In the other limit of strong couplings,
$1/G$ -- proportional to the resistivity $\rho$ -- decreases with increasing temperature, indicating 
a clear insulating phase, as a
result of the system being in a clustered state at the temperatures 
explored,\cite{paper0,paper1} with carriers localized near the Mn spins. 
At the important intermediate couplings emphasized in our effort, 
the system behaves like a dirty metal for $T<T_{\rm C}$, while for
$T_{\rm C}<T<T^*$, $1/G$ slightly decreases with increasing temperature, 
indicating that a soft metal to
insulator transition takes place near $T_{\rm C}$.  
For $T>T^*$, where the system is paramagnetic, $1/G$ is almost constant. 
Note that for strong enough $J/t$, $T_{\rm C}\rightarrow0$ 
and therefore no metallic phase is present. 

It is interesting to compare these numerical results with experiments. 
For Ga$_{1-x}$Mn$_x$As, data similar to those found in our
investigations have been reported.\cite{katsumoto,ohno1,potashnik}
The qualitative behavior of the resistivity in these 
samples agrees well with the theoretical results presented 
in Fig.~\ref{fig:rho4d3}b if intermediate couplings are considered. 

The clustered state that forms just above $T_{\rm C}$ is a candidate
to describe DMS materials since it explains both the resistivity maximum around
$T_{\rm C}$, as well as the decrease in resistivity with increasing applied 
magnetic field. In addition, it provides an optimal $T_{\rm C}$. This clustered
 state could be unveiled using scanning tunneling microscopy techniques. 

Recent experimental work on (Ga,Cr)As have revealed unusual magnetic
properties which were associated with the random magnetism of the alloy.
The authors of Ref.~\cite{dakhama} explained their results using a distributed
magnetic polaron model, that resembles the clustered-state ideas
discussed here and in Refs.~\cite{paper1,timm}. The present simulations unveils a
\emph{previously unknown similarity between DMS materials and transition metal oxides}, 
since both have regimes where clustered states dominate. 
This analogy deserves further work.

\section{Conclusions}

The information provided here is indicative of the key role
that inhomogeneous clustered states play in many materials. Their existence
leads to quite nontrivial results. The colossal magnetoresistance
effect appears to be linked to their presence. Giant proximity effects
in cuprates may have a similar origin. The physics of important materials
for applications such as diluted magnetic semiconductors may share a
similar phenomenology, including a large magnetoresistance. This field
is at an early stage in its development, which will need a coordinated
effort between theorists -- mainly those with expertise in simulations --
and experimentalists. The study of clustered states is rapidly developing
into one of the most active areas of present day condensed matter physics.

Work supported by NSF grant DMR-0122523.

\end{document}